\documentclass[10pt, article,twocolumn]{IEEEtran}
\usepackage{times}
\usepackage[cmex10]{amsmath}
\usepackage{amssymb}
\usepackage{epsfig,verbatim}
\usepackage{algorithm}
\usepackage{algpseudocode}

\usepackage{graphicx,color,epsfig,rotating,subfigure}
\usepackage{amsfonts,amsmath,amssymb}
\usepackage{algorithm}
\usepackage{subfigure}
\usepackage{algpseudocode}

\newcommand\blfootnote[1]{%
  \begingroup
  \renewcommand\thefootnote{}\footnote{#1}%
  \addtocounter{footnote}{-1}%
  \endgroup
}

\long\def\comment#1{}

\newfont{\bbb}{msbm10 scaled 700}

\newfont{\bb}{msbm10 scaled 1100}
\newcommand{\CC}{\mbox{\bb C}}
\newcommand{\PP}{\mbox{\bb P}}
\newcommand{\RR}{\mbox{\bb R}}

\newcommand{\av}{{\bf a}}
\newcommand{\bv}{{\bf b}}
\newcommand{\cv}{{\bf c}}

\newcommand{\hv}{{\bf h}}

\newcommand{\qv}{{\bf q}}
\newcommand{\rv}{{\bf r}}

\newcommand{\uv}{{\bf u}}
\newcommand{\wv}{{\bf w}}

\newcommand{\xv}{{\bf x}}
\newcommand{\yv}{{\bf y}}
\newcommand{\zv}{{\bf z}}

\newcommand{\Hm}{{\bf H}}

\newcommand{\Ac}{{\cal A}}

\newcommand{\Cc}{{\cal C}}

\newcommand{\Ec}{{\cal E}}

\newcommand{\Ic}{{\cal I}}

\newcommand{\Nc}{{\cal N}}

\newcommand{\Sc}{{\cal S}}

\newcommand{\Wc}{{\cal W}}

\newcommand{\alphav}{\hbox{\boldmath$\alpha$}}
\newcommand{\betav}{\hbox{\boldmath$\beta$}}

\newcommand{\muv}{\hbox{\boldmath$\mu$}}

\newcommand{\SNR}{{\sf SNR}}

\newcommand{\eqdef}{\stackrel{\Delta}{=}}

\newcommand{\transp}{{\sf T}}

\newtheorem{definition}{Definition}

\newcommand{\argmin}{\operatornamewithlimits{argmin}}

\newtheorem{example}{Example}

\title{A  Low-Complexity Soft-Output wMD Decoding for Uplink MIMO Systems with One-Bit ADCs}

\author{
\IEEEauthorblockN{
              Seonho Kim\authorrefmark{1},  Namyoon Lee\authorrefmark{2}, and Song-Nam Hong\authorrefmark{1}}\\
\IEEEauthorblockA{\authorrefmark{1}Ajou University, Suwon, Korea,\\
              email: \{shkim1005, snhong\}@ajou.ac.kr}\\
\IEEEauthorblockA{\authorrefmark{2}POSTECH, Pohang, Korea,\\
              email: nylee@postech.ac.kr}
}

\begin{document}

\maketitle

\date{}

\blfootnote{
}

\begin{abstract}
This paper considers an uplink multiuser multiple-input-multiple-output (MU-MIMO) system with one-bit analog-to-digital converters (ADCs), in which $K$ users with a single transmit antenna communicate with one base station (BS) with $N_{\rm r}$ receive antennas. In this system, a novel MU-MIMO detection method, named {\em weighted minimum distance} (wMD) decoding, was recently proposed, as a practical approximation of maximum likelihood (ML) detector. Despite of its attractive performance, the wMD decoding has two limitations to be used in practice: i) the hard-decision outputs degrade the performance of a following channel code; ii) the computational complexity grows exponentially with the $K$. To address them, we first present a {\em soft-output} wMD decoding that efficiently computes soft metrics (i.e., log-likelihood ratios) from one-bit quantized observations. We then reduce the complexity of the soft-output wMD decoding by introducing {\em hierarchical code partitioning}. Simulation results demonstrate that the proposed method significantly outperforms the other MIMO detectors with a comparable complexity.
\end{abstract}
\begin{keywords}
Multiuser MIMO detection, analog-to-digital converter (ADC), one-bit ADC.
\end{keywords}

\section{Introduction}


The use of a very large number of antennas at the base station (BS), referred to as massive multiple-input-multiple-output (MIMO), is one of the promising techniques to cope with the predicted wireless data traffic explosion \cite{Marzetta}-\cite{Lu}. The massive MIMO can improve the system throughput and energy efficiency  \cite{Lu, Yang}. In contrast, it can considerably increase the hardware cost and the radio-frequency (RF) circuit consumption \cite{Yang}. Among all the components in a RF chain, a high-resolution analog-to-digital converter (ADC) is particularly power-hungry as the power consumption of an ADC is scaled exponentially with the number of quantization bits and linearly with the baseband bandwidth  \cite{Murmann, Mezghani-2011}. To overcome this challenge, the use of low-resolution ADCs  (e.g., 1$\sim$3 bits) for massive MIMO systems has received increasing attention over the past years. The one-bit ADC is particularly attractive because of the lower hardware complexity. In this case, the in-phase and quadrature components of the continuous-valued received signals are quantized separately using simple zero-threshold comparators and there is no need for an automatic gain controller \cite{Hoyos}. Despite the benefits of using low-resolution ADCs, it gives rise to numerous technical challenges: i) an accurate channel estimation at the receiver (CSIR) is complicated; ii) conventional MIMO detection methods, developed for linear MIMO systems, yield a poor bit error rates (BERs) as the impact of non-linearity of ADCs was not taken into account.


There have been extensive works on the MIMO detection and channel estimation methods for the uplink MIMO systems with one-bit ADCs \cite{Risi}-\cite{Hong_J}. The optimal maximum likelihood (ML) detection was introduced in \cite{Choi} and low-complexity methods were also presented in  \cite{Choi, Mollen, Mollen2}. Also, numerous channel estimation methods using  one-bit quantized observations were developed as least-square (LS) based method \cite{Risi}, maximum-likelihood (ML) type method \cite{Choi}, zero-forcing (ZF) type method \cite{Choi}, and Bussgang decomposition based method \cite{Li}.  Recently, a novel MIMO detection method, named {\em weighted minimum distance} (wMD) decoding, was presented by viewing the MIMO detection problem as an equivalent coding problem \cite{Hong_J}. The equivalent coding problem is to find a codeword of the spatial-domain code $\Cc$ from the one-bit quantized observations obtained from $2N_{\rm r}$ parallel channels with unequal channel reliabilities (see Fig.~\ref{e_model}), where the code $\Cc$ is not designable but is completely determined as a function of a channel matrix $\Hm$. In this problem,  the wMD decoding, as an extension of minimum distance (MD) decoding, was presented by exploiting the distinct channel reliabilities appropriately. Furthermore, it was demonstrated that the wMD decoding achieves the optimal ML performance for a perfect CSIR and is more robust to an inaccurate CSIR than ML detector \cite{Hong_J}.

\begin{figure*}
\centerline{\includegraphics[width=16cm]{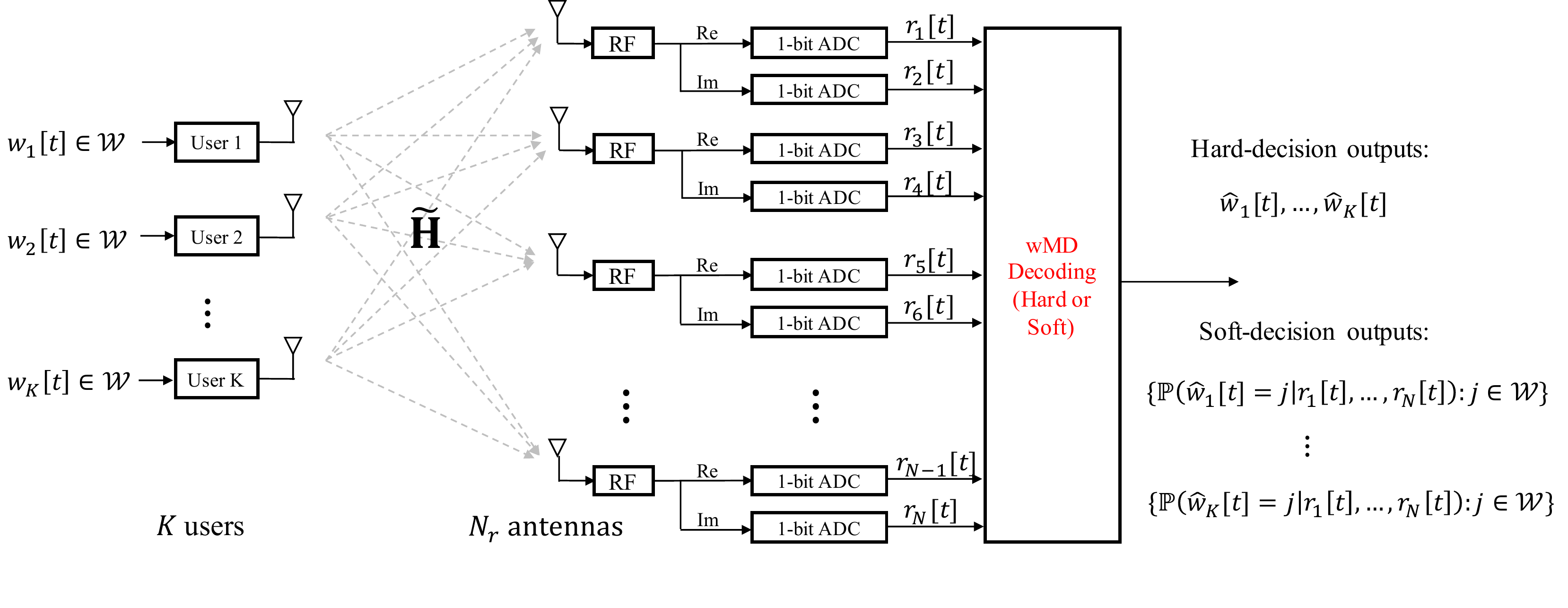}}
\caption{Uplink  MU-MIMO systems in which each receive antenna at a BS is equipped with one-bit ADCs.}
\label{model}
\end{figure*}


Despite of its attractive performance, there are two technical challenges so that the wMD decoding will be adopted in practical communication systems. First, the wMD decoding produces the hard-decision outputs as in the other MIMO detection methods in \cite{Choi,Lee}, which degrades the performance of a following channel decoder. Also, the computational complexity is not manageable when the number of active users is large. In this paper, we address the above problems, by presenting a soft-output wMD decoding and by reducing its complexity using hierarchical code partitioning. Our contributions are summarized as follows.


\begin{itemize}


\item We propose a {\em soft-output} wMD decoding for the uplink MU-MIMO systems with one-bit ADCs. The proposed soft-output wMD decoding produces the soft outputs (e.g., log-likelihood ratios (LLRs)) from one-bit quantizized (hard-decision) observations. 
This enables to employ a state-of-the-art soft channel decoder (e.g., belief-propagation decoder \cite{Richardson}). Whereas, the previous MIMO detection methods in \cite{Choi, Lee, Hong_J}  produces the hard-decision outputs and hence, a highly suboptimal hard channel decoder (e.g., bit-flipping decoder \cite{Rao}) should be used.


\item We reduce the complexity of the soft-output wMD decoding using the idea of a sphere decoding, in which some unnecessary codewords of the $\Cc$ are precluded from the search-space.  The key idea is to partition the spatial-domain code $\Cc$ (i.e., the overall search-space) into the several subcodes in a {\em hierarchical} manner: the $\Cc$ is partitioned into the level-1 subcodes and then each level-1 subcode is further partitioned into the level-2 subcodes, and so on (see Fig.~\ref{multilevel}). This process is referred to as {\em hierarchical code partitioning}. Leveraging this structure, we can efficiently define a reduced code $\Cc_{\rm r}(\rv[t])$ only containing the codewords of the $\Cc$ close to the current observations $\rv[t]$.

\item Simulation results demonstrate that the proposed MIMO detection method significantly outperforms the other MIMO detection methods with a comparable complexity. It is remarkable that the performance gain is essentially attained by the soft outputs obtained from one-bit quantized observations.


\end{itemize}

The outline of this paper is as follows. In Section~\ref{sec:pre}, we describe the system model of uplink MIMO system with one-bit ADCs and review the wMD decoding. In Section~\ref{sec:s-wMD}, we present a soft-output wMD decoding which efficiently computes soft metrics from one-bit quantized observations.  In Section~\ref{sec:main}, a low-complexity (soft-output) wMD decoding is presented by introducing hierarchical code partitioning. Section~\ref{sec:numerical} provides the numerical results to show the superiority of the proposed method. Section~\ref{sec:conclusion} concludes the paper.

{\bf Notation:} Lower and upper boldface letters represent column vectors and matrices, respectively.  For any $k \in \{0,...,K-1\}$, we let $g(k)=[b_0,b_1,\ldots,b_{K-1}]^{\transp}$ represent the $m$-ary expansion of $k$ where 
$k=b_0m^0+\cdots+b_{K-1}m^{K-1}$ for $b_i \in \{0,...,m-1\}$.
 We also let $g^{-1}(\cdot)$ denote its inverse function. For a vector, $g(\cdot)$ is applied element-wise. Likewise, if a scalar function is applied to a vector, it will be performed element-wise. ${\rm Re}(\av)$ and ${\rm Im}(\av)$ represent the real and complex part of a complex vector $\av$, respectively.

\section{Preliminaries} \label{sec:pre}

In this section, we define an uplink multiuser MIMO system with one-bit ADCs and review the wMD decoding proposed in~\cite{Hong_J}.

\subsection{System Model}\label{sec:model}\label{sec:SM}

\begin{figure}[t]
\centerline{\includegraphics[width=8cm]{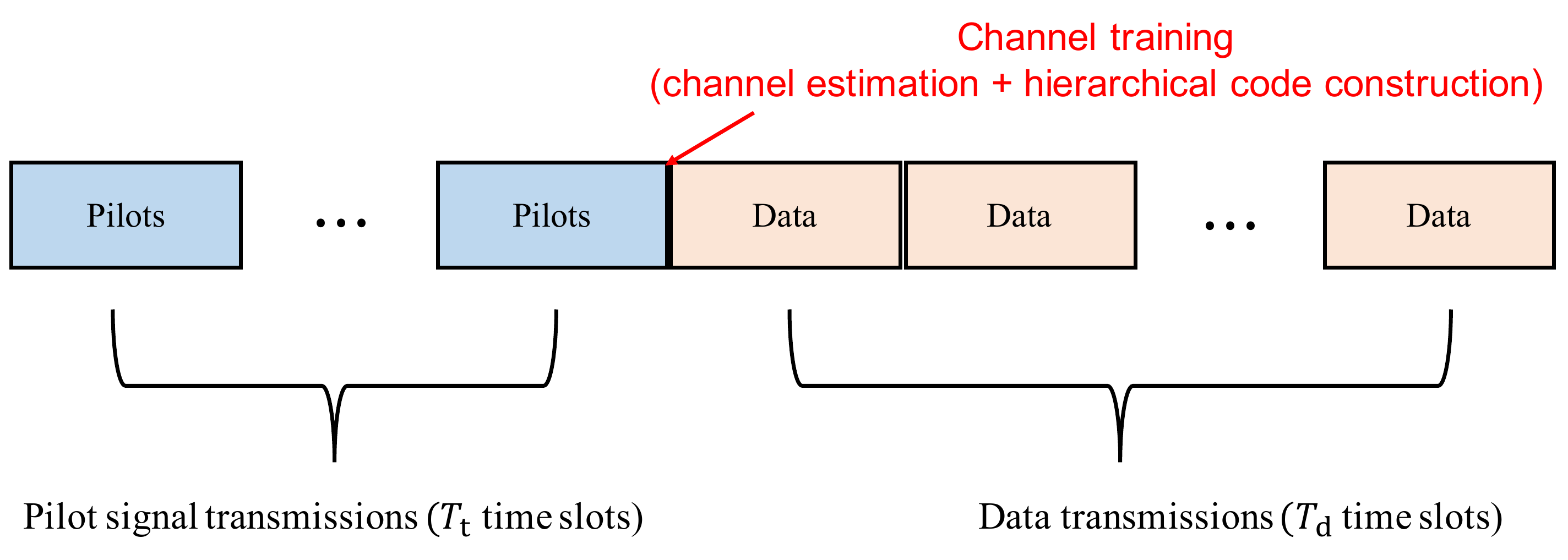}}
\caption{Frame structure consisting of the channel training and data transmissions, during a coherence time.}
\label{channel_training}
\end{figure}

We consider a single-cell uplink multiuser MIMO system in which $K$ users with a single-antenna communicate with one  BS with an array of $N_{\rm r} > K$ antennas (see Fig.~\ref{model}). We use the $t$ to indicate a time-index. Let $w_k[t] \in \mathcal{W}=\{0,...,m-1\}$ represent the user $k$'s message for $k \in\{1,...,K\}$, each of which contains $\log{m}$ information bits. We also denote $m$-ary constellation set by $\Sc=\{s_0,...,s_{m-1}\}$ with power constraint as
\begin{equation}
\frac{1}{m}\sum_{i=0}^{m-1}\|s_i\|^2 = \SNR.
\end{equation}
Then, the transmitted symbol of user $k$ at time $t$, ${\tilde x}_k(w_k[t])$, is obtained by a modulation function $f:\mathcal{W}\rightarrow \Sc$ as
\begin{equation}
\tilde{x}_k(w_k[t]) = f(w_k[t]) \in \Sc.
\end{equation} 
When the $K$ users transmit the symbols ${\tilde \xv}(\wv[t])=[\tilde{x}_1(w_1[t]),\ldots,\tilde{x}_K(w_K[t])]^{\transp}$, the discrete-time complex-valued baseband received signal vector at the BS, ${\bf \tilde r}[t]\in\mathbb{C}^{N_{\rm r}}$, is given by
\begin{equation}
{\bf \tilde r}[t] = {\bf \tilde H}{\bf \tilde x}(\wv[t]) +{\bf \tilde z}[t], \label{eq:system_complex}
\end{equation} where ${\bf \tilde H} \in \CC^{N_{\rm r} \times K}$ is the channel matrix between the BS and the $K$ users, i.e., the $i$-th row of ${\bf \tilde H}$ is the channel vector between the $i$-th receive antenna at the BS and the $K$ users.   In addition, ${\bf \tilde z}[t]=[{\tilde z}_1[t],\ldots,{\tilde z}_{N_{\rm r}}[t]]^{\transp}\in\mathbb{C}^{N_{\rm r}}$ is the noise vector whose elements are distributed as circularly symmetric complex Gaussian random variables with zero-mean and unit-variance, i.e., ${\tilde z}_i[t] \sim \Cc\Nc(0,1)$. We assume a block fading channel in which the channel matrix $\Hm$ remains constant during $T$ time slots (e.g., coherence time).  A transmission frame containing $T_{\rm c}$ time slots is composed of two different types of a frame as a pilot transmission frame and a data transmission frame (see Fig.~\ref{channel_training}). The first $T_{\rm t}$ time slots are allocated for the pilot transmission frame and the subsequent $T_{\rm d}$ time slots are allocated for the data transmission frame, i.e., $T_{\rm c}=T_{\rm t} + T_{\rm d}$. During the pilot transmission frame, the $K$ users send the pilot signals that are known at the BS, while during the data transmission frame, the users send the data signals that convey the information to the BS.

In the MIMO system with one-bit ADCs,  each receive antenna of the BS is equipped with RF chain followed by two one-bit ADCs that are applied to each real and imaginary part separately.  Let $\mbox{sign}(\cdot): \RR \rightarrow \{0,1\}$ represent the one-bit ADC quantizer function with
\begin{equation}
\hat{r}[t]=\mbox{sign}(\tilde{r}[t])=\begin{cases}
0 & \mbox{ if } \tilde{r}[t] \geq 0\\
1 & \mbox{ if } \tilde{r}[t]  < 0.
\end{cases}
\end{equation}  Then, the BS receives the quantized output vector as
\begin{align}
\hat{\rv}_{\rm R}[t] = \mbox{sign}({\rm Re}({\bf \tilde r}[t]))\mbox{ and }\hat{\rv}_{\rm I} =\mbox{sign}({\rm Im}({\bf \tilde r}[t])).
\end{align} 
For the ease of representation, we rewrite the complex input-output relationship in \eqref{eq:system_complex} into the equivalent real representation as
\begin{equation}
\rv[t] = \mbox{sign}\left(\Hm\xv(\wv[t])+\zv[t]\right), \label{eq:obs1}
\end{equation}
where $\rv[t]=[\hat{\rv}_{\rm R}[t]^{\transp},\hat{\rv}_{\rm I}[t]^{\transp}]^{\transp}$, $\xv(\wv[t])=[\mbox{Re}(\tilde{\xv}(\wv[t]))^{\transp},\mbox{Im}(\tilde{\xv}(\wv[t]))^{\transp}]^{\transp}$, $\zv[t]=[\mbox{Re}(\tilde{\zv}[t])^{\transp},\mbox{Im}(\tilde{\zv}[t])^{\transp}]^{\transp}$, and 
\begin{equation*}
\Hm = \left[ {\begin{array}{cc}
   \mbox{Re}({\bf \tilde H}) & -\mbox{Im}({\bf \tilde H}) \\      
   \mbox{Im}({\bf \tilde H}) & \mbox{Re}({\bf \tilde H})\\
 \end{array} } \right] \in\mathbb{R}^{N\times 2K},
 \end{equation*}
and where $N=2N_{\rm r}$. This real system representation will be used in the sequel.

\subsection{wMD Decoding}\label{sec:wMD}

\begin{figure}[t]
\centerline{\includegraphics[width=9cm]{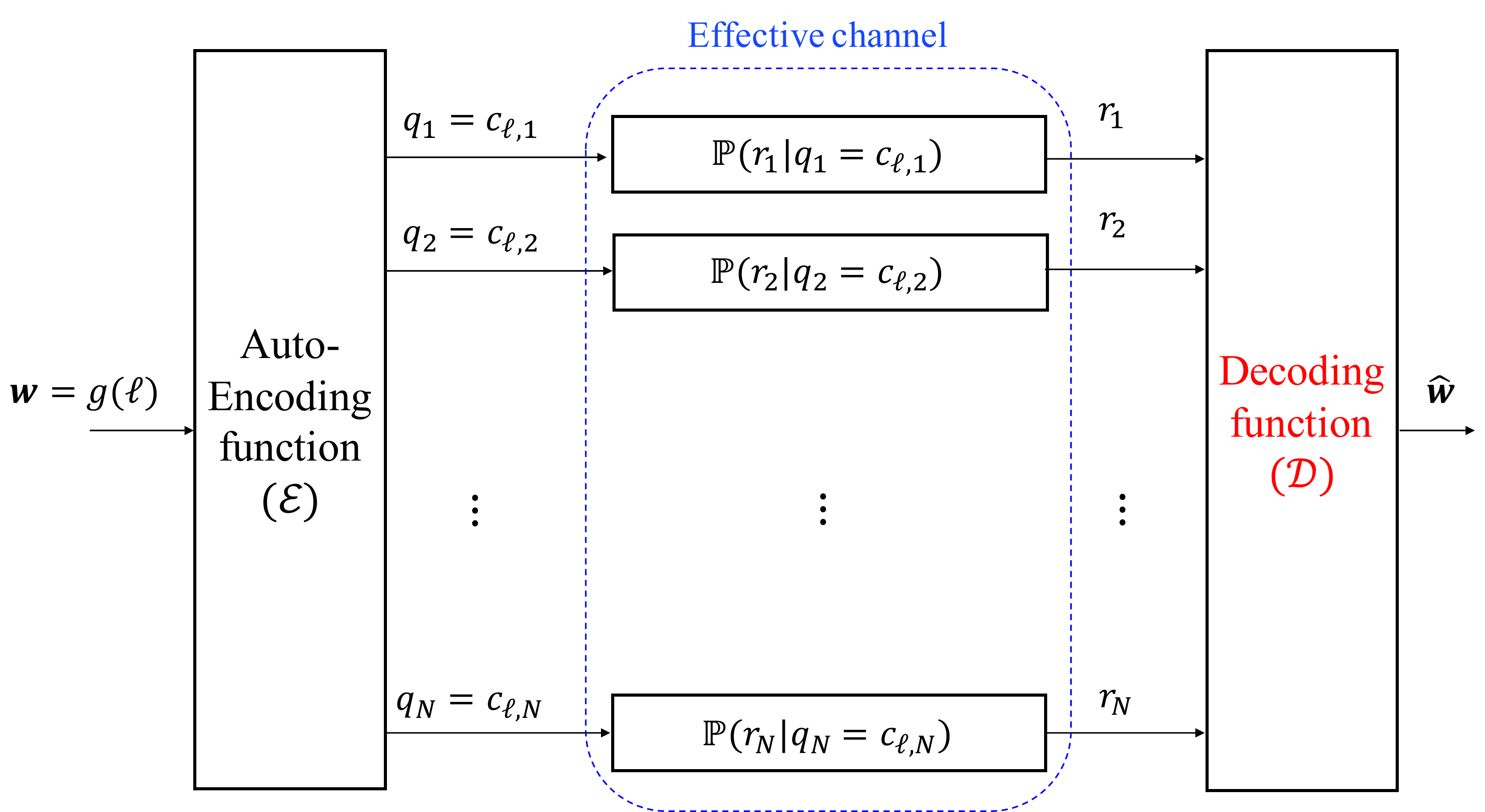}}
\caption{Description of an equivalent coding problem. Note that an auto-encoding function $\Ec$ is determined as a function of $\Hm$ and a one-bit quantization function. Also, the transition probabilities of an effective channel depend on the message vector $\wv$ (i.e., asymmetric channel).}
\label{e_model}
\end{figure}

We review the wMD decoding presented in \cite{Hong_J}. This method was developed by showing the equivalence of the original MIMO detection problem and a non-linear coding problem (see Fig.~\ref{e_model}). The equivalent coding problem consists of the three parts as described below. Since this method is applied symbol-by-symbol, we in this section drop the time-index $t$ for the ease of exposition. It is assumed that, during the channel training phase, a channel matrix $\Hm$ is estimated at the BS. Then, we will explain the wMD decoding which is performed to decode the users' messages during the data transmission phase.

\vspace{0.1cm}
{\em i) Auto-encoding function:}  For a given channel matrix $\Hm$, a code $\Cc$ over a spatial domain is defined as
\begin{equation}\label{eq:code-def}
\Cc=\{\cv_0,\ldots,\cv_{m^K-1}\},
\end{equation} where each codeword $\cv_{\ell}$ is defined as
\begin{equation*}
\cv_\ell= \left[\mbox{sign}\left(\hv_{1}^{\transp}\xv(g(\ell))\right),\ldots, \mbox{sign}\left(\hv_{N}^{\transp}\xv(g(\ell))\right)\right]^{\transp}.
\end{equation*} The code $\Cc$ is a non-linear binary code of length $N$ and code rate $\frac{K\log{m}}{N}$. Since this code is completely described as a function of channel matrix $\Hm=[\hv_1,\ldots,\hv_{N}]^{\transp}$, this code is referred to as a {\em spatial-domain} code. Also, we call a channel code {\em time-domain} code.

In Fig.~\ref{e_model}, the input $\qv[t]$ of an effective channel is generated by an auto-encoding function $\Ec :\{0,...,m-1\}^K \rightarrow \Cc$ as
\begin{equation}
\qv=\Ec(\wv) = \cv_{\ell}
\end{equation}  where $\ell = g^{-1}(\wv) \in \{0,1,...,m^K-1\}$.\\

\begin{example} Consider a $2\times 2$ MIMO system with one-bit ADC, and each user is assumed to use 4-QAM, i.e.,  $N_{\rm r}=2$, $K=2$, and $m=4$. Then, for a given channel matrix ${\bf H}\in\mathbb{R}^{4\times 4}$, one can create a code $\mathcal{C}=\{{\bf c}_1,{\bf c}_2,\ldots, {\bf c}_{16}\}$ in which the $\ell$-th codeword is defined as
\begin{equation*}
{\bf c}_\ell=\left[\mbox{sign}\left(\hv_{1}^{\transp}\xv(g(\ell))\right),\ldots, \mbox{sign}\left(\hv_{4}^{\transp}\xv(g(\ell))\right)\right]^{\transp}\in\{0,1\}^4.
\end{equation*}
\end{example}

{\em ii) Effective channel:}
As shown in Fig.~\ref{e_model}, the effective channel consists of $N$ parallel binary input/output channels with input $\qv=[q_1,\ldots,q_{N}]^{\transp}$ and output $\rv=[r_1,\ldots,r_{N}]^{\transp}$. For the $i$-th subchannel, the transition probabilities, depending on users' messages $\wv=g(\ell)$,  are defined as
\begin{equation}
p_{\ell,i,j}\eqdef\PP(r_i= j | q_i=c_{\ell,i}),\label{eq:trans_prob}
\end{equation} for $j \in \{0,1\}$. This is simply computed using Q-function as
\begin{equation}
p_{\ell,i,j} = \begin{cases}
\epsilon_{\ell,i} & \mbox{ if } i\neq j\\
1 - \epsilon_{\ell,i} & \mbox{ if } i=j.
\end{cases}
\end{equation}where $\epsilon_{\ell,i} \eqdef Q(|\hv_{i}^{\transp}\xv(g(\ell))|<0)$ denotes a cross-probability of the channel $i$ and 
\begin{equation*}
Q(x) = \frac{1}{2\pi}\int_{x}^{\infty} \exp\left(-\frac{x^2}{2}\right) dt.
\end{equation*} 

\vspace{0.1cm}
{\em iii) Decoding function:} The wMD decoding was presented in \cite{Hong_J} as an extension of a minimum distance (MD) decoding.
\begin{definition} \label{def:wMD} A {\em weighted} Hamming distance is defined as
\begin{equation*}
d_{\rm wh}(\xv,\yv; \alphav) \eqdef \sum_{i=1}^{N} \alpha_i \mathbf{1}_{\{x_i \neq y_i\}},
\end{equation*} where $\alphav=(\alpha_1,...,\alpha_N)$ denotes a weight vector, $\mathbf{1}_{\Ac}$ represents an indicator function with $\mathbf{1}_{\Ac}=1$ if $\Ac$ is true, and $\mathbf{1}_{\Ac}=0$, otherwise. Note that the Hamming distance is a special case of the weighted Hamming distance with equal weights (i.e., $\alpha_i = 1$ for all $i$). 
\end{definition}
\vspace{0.1cm}

Using the definition, the wMD decoding is performed as
\begin{equation}
\hat{\ell} = \argmin_{\ell \in [1:m^K]} d_{\rm wh} (\rv, \cv_{\ell}; \alphav_{\ell}),\label{eq:wMDD}
\end{equation} where the weights are defined using the channel reliabilities as
\begin{equation}
\alpha_{\ell,i} = -\log\left(Q(|\hv_{i}^{\transp}\xv(g(\ell))|<0)\right),
\end{equation} for $i \in \{0,1...,m^K-1\}$. The key idea of the wMD decoding is to allocate a higher belief to the information conveyed from a more reliable channel while MD decoding assigns an identical belief. Also, it was demonstrated in \cite{Hong_J} that the wMD decoding outperforms MD decoding due to the use of the weights.

\section{Soft-Output wMD Decoding}\label{sec:s-wMD}

Likewise ML and near ML detectors in \cite{Choi}, and supervised-learning based detector in \cite{Lee}, the wMD decoding produces the hard-decision outputs. Inevitably, a hard channel decoder (e.g., bit-flipping decoder) should be employed as in \cite{Choi}. This approach can yield a non-trivial performance loss compared to using soft channel decoder (e.g., belief-propagation decoder). To overcome this problem, we propose a soft-output wMD decoding which generates a soft metric (e.g., LLR) from one-bit quantized (hard-decision) observation.

We first define the subcode of the $\Cc$ as follows:

\begin{definition} \label{def:subcode} Recall that a spatial-domain code $\Cc$ is defined as
\begin{equation}
\Cc\eqdef\{\cv= \Ec(\wv): \wv \in \Wc^{K}\}.
\end{equation} For any given user's message $\{w_{k} = j\}$ with $j \in \Wc$, the subcode of the $\Cc$ is defined as
\begin{align*}
\Cc_{|\{w_{k} = j\}} \eqdef \{\cv= \Ec(\wv): \wv \in \Wc^{K}, w_{k} = j\}.
\end{align*}
\end{definition}
\vspace{0.2cm}

Using the above definition, we will compute the a posteriori probabilities (APPs) from the one-bit quantized observation $\rv[t]=(r_1[t],...,r_N[t])$, where the APPs are defined as

\begin{equation}
\left\{\PP\{w_k[t] = j |\rv[t]\}: j \in \Wc, k \in \{1,...,K\}\right\}. \label{eq:APPs}
\end{equation}  We let $\wv_{\bar{k}}[t] = (w_1[t],..,w_{k-1}[t]$ $,w_{k+1}[t],..,w_{K}[t])^T$. Then, the APP of the user $k$'s message is computed as

\begin{align}
\PP (w_k [t]= j |\rv[t])&=\sum_{\uv \in \Wc^{K-1}}\PP(w_k [t]= j, \wv_{\bar{k}}[t] = \uv |\rv[t])\nonumber\\
& \stackrel{(a)}{=}\frac{1}{Z}\sum_{\uv\in\Wc^{K-1}} \PP(\rv[t] |w_k[t] = j, \wv_{\bar{k}}[t]=\uv)\nonumber\\
& \stackrel{(b)}{=}\frac{1}{Z}\sum_{\cv_{\ell} \in \Cc_{|w_k[t]=j} }\PP(\rv[t]|\cv_{\ell}),\label{eq:APP}
\end{align} for $j \in \Wc$, where (a) is from the Bayes' rule, (b) is from Definition~\ref{def:subcode},  $\PP(\rv[t]|\cv_{\ell})$ is defined in (\ref{eq:trans_prob}), and $Z$ denotes a normalization factor such that
\begin{equation}
\sum_{j \in \Wc} \PP (w_k [t]= j |\rv[t])=1.
\end{equation}  Using the weighted Hamming distance in Definition 1, the (\ref{eq:APP}) can be approximately computed as
\begin{equation}\label{eq:APP-1}
\PP (w_k[t] = j | \rv[t]) \approx \frac{1}{Z} {\rm exp}\left(- \sum_{ \cv_{\ell} \in \Cc_{|w_k [t]=j} } d_{{\rm wh}}(\rv[t], \cv_{\ell} , \alphav_{\ell})\right).
\end{equation} Note that the above approximation is very accurate when the crossover probability of each subchannel is smaller than 0.3 \cite{Hong_J}.
Also, using the well-known approximation  as
\begin{equation}\label{approx}
{\rm exp}(x_1+x_2+\cdots+x_t) \approx {\rm exp} (\max\{x_1,x_2,...,x_t\}),
\end{equation} the (\ref{eq:APP-1}) can be further simplified as
\begin{equation}
\PP (w_k [t] = j | \rv[t]) \approx \frac{1}{Z} {\rm exp} \left(- \min_{ \cv_{\ell} \in \Cc_{|w_k[t]=j} } d_{{\rm wh}}(\rv[t],\cv_{\ell}.\alphav_{\ell}\})\right),\label{prob-app}
\end{equation}

\begin{figure*}
\centerline{\includegraphics[width=16cm]{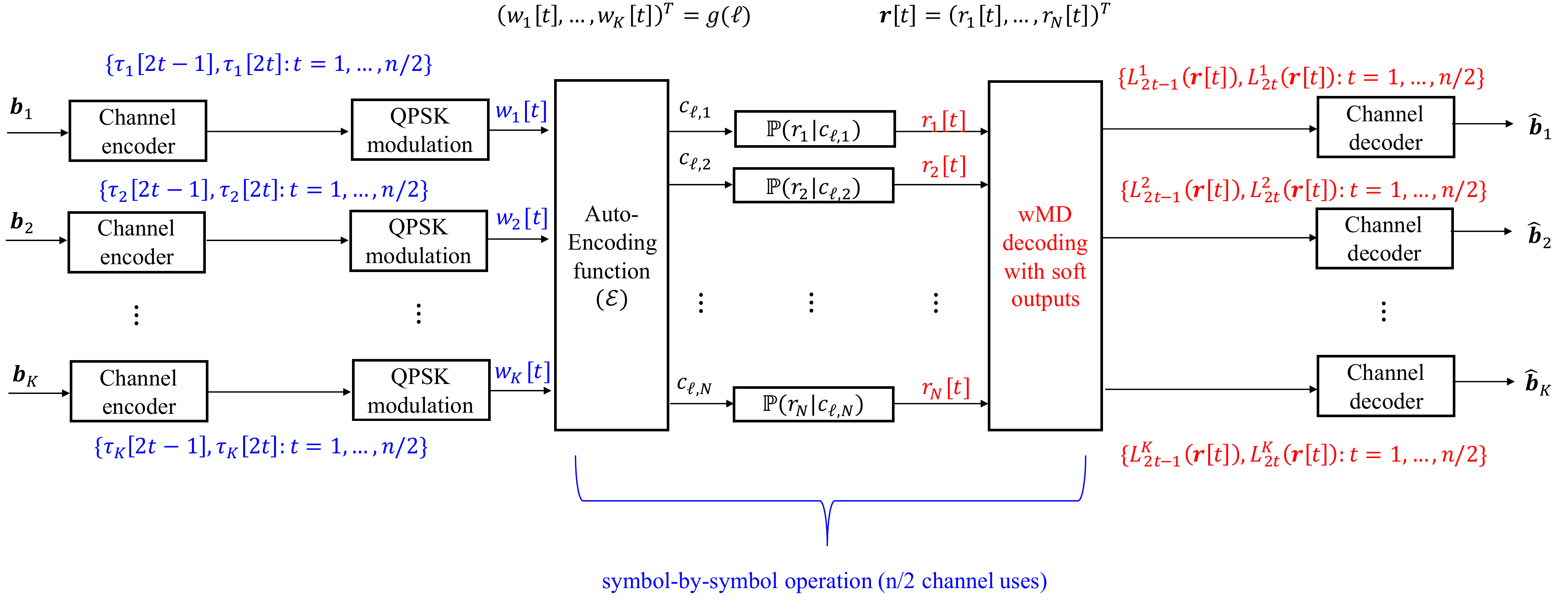}}
\caption{The proposed coded architecture for uplink MU-MIMO systems with one-bit ADCs.}
\label{coded_EM}
\end{figure*}


From the APPs derived in  (\ref{prob-app}) (or (\ref{eq:APPs})), we then compute the soft inputs (e.g., log-likelihood ratios (LLRs) ) of a channel decoder. To make an explanation clear, we only consider a $2^q$-QAM constellation (e.g., $\Wc=\{0,1,...,2^q\}$ for some positive $q$. However, the extension to an arbitrary $m$-ary constellation is straightforward. Fig.~\ref{coded_EM} describes the coded system for $q=2$ (i.e., 4-QAM).  Let $(\tau_k[1],...\tau_k[n])$ dente the coded output of the user $k$'s channel encoder. For the ease of notation, we define:
\begin{equation}
[\bv]_{q} \eqdef \sum_{i=1}^{q} b_i 2^{q-i},
\end{equation}where $\bv=(b_1,...,b_q)$ with $b_{i} \in \{0,1\}$ for $i=1,..,q$. Then, the user $k$'s channel input message at time slot $t$ is obtained as
\begin{equation}
w_k[t]  = [(\tau_{k}[qt],\tau_{k}[qt-1],...,\tau_{k}[qt-q+1])]_{q},
\end{equation} for $t=1,...,n/q$, where it is assumed that $n$ is a multiple of $q$. Each user $k$ transmits the $\{w_k[t]: t=1,..,n/q\}$ to the BS over the $n/q$ time slots. 
From the observations $\{\rv[t]: t=1,...,n/2\}$ and using  (\ref{prob-app}), the BS first computes the APPs as
\begin{equation}
\{\PP (w_k[t] = j | \rv[t]): j \in \Wc, t=1,...,n/q\}.\label{eq:probs-1}
\end{equation} Then, it computes the soft inputs (e.g., LLRs) of the channel decoder as
\begin{align*}
L_{qt-(i-1)}^{k}(\rv[t]) &\eqdef \log{\frac{\PP (\tau_k [qt-(i-1)]= 0| \rv[t])}{\PP (\tau_k [qt-(i-1)] = 1| \rv[t])}}\\
&= \log\frac{\sum_{\bv \in\{0,1\}^q:b_i = 0} \PP(w_{k}[t] = [\bv]_{q})  }{\sum_{\bv\in\{0,1\}^q:b_i = 1} \PP(w_{k}[t] = [\bv]_{q}) },
\end{align*} for $i=1,...,q$ and  $t=1,...,n/q$. This can be simply computed from (\ref{approx}) and (\ref{prob-app}) as
\begin{align}
L_{qt-(i-1)} &= \min_{ \cv_{\ell} \in \bigcup_{\bv \in \{0,1\}^q :b_i = 1} \Cc_{|w_k[t]= [\bv]_{q}}} d_{{\rm wh}}(\rv[t],\cv_{\ell} , \alphav_{\ell})\nonumber\\
& -\min_{ \cv_{\ell} \in \bigcup_{\bv \in \{0,1\}^q :b_i = 0} \Cc_{|w_k[t]= [\bv]_{q}}} d_{{\rm wh}}(\rv[t],\cv_{\ell} , \alphav_{\ell}),\label{eq:LLRs}
\end{align} for $i=1,...,q$ and $t=1,...,n/q$.

\begin{example} When 4-QAM is used, the LLRs are computed from (\ref{eq:LLRs}) as 
\begin{align*}
 L_{2t-1}^{k}(\rv[t])=&\min_{\cv_{\ell} \in  \Cc_{|w_k[t]=2}\; \cup \; \Cc_{|w_k[t]=3}} d_{{\rm wh}}(\rv[t],\cv_{\ell} , \alphav_{\ell})\\
 & - \min_{\cv_{\ell} \in  \Cc_{|w_k[t]=0}\;  \cup\;   \Cc_{|w_k[t]=1}} d_{{\rm wh}}(\rv[t],\cv_{\ell} ,\alphav_{\ell})\\
L_{2t}^{k}(\rv[t])= &\min_{\cv_{\ell} \in  \Cc_{|w_k[t]=1}\; \cup \; \Cc_{|w_k[t]=3}} d_{{\rm wh}}(\rv[t],\cv_{\ell} , \alphav_{\ell})\\
&- \min_{\cv_{\ell} \in  \Cc_{|w_k[t]=0}\; \cup \; \Cc_{|w_k[t]=2}} d_{{\rm wh}}(\rv[t],\cv_{\ell} , \alphav_{\ell}),
\end{align*}for $t=1,...,n/2$.
\end{example}

\section{A Low-Complexity Soft-Output wMD Decoding \\Using Hierarchical Code Partitioning}\label{sec:main}

We observe that the computational complexity of the soft-output wMD decoding as well as wMD decoding is problematic for a large $K$ as the size of the code $\Cc$ (i.e., the search-space) grows exponentially with the $K$. In this section, we present a low-complexity soft-output wMD decoding by introducing hierarchical code structure. Note that the proposed method is directly applied to the wMD decoding. The key idea of the proposed method is that the code $\Cc$ is partitioned in a hierarchical manner:  the code $\Cc$ is partitioned into the level-1 subcodes and each level-1 subcode is further partitioned into the level-2 subcodes, and so on (see Section~\ref{subsec:CT}). This process is referred to as {\em hierarchical code partitioning}. Leveraging this structure, we can efficiently identify some codewords of the $\Cc$ that lie inside the sphere centered at the current observation $\rv[t]$ with a certain radius, where the reduced code is denoted by $\Cc_{\rm r}(\rv[t])$. This method is reminiscent of a sphere decoding \cite{Schnorr, Fincke} that is developed for conventional MIMO systems. In this sense, the proposed method can be regarded as a sphere decoding for the MIMO systems with one-bit ADCs.

To be specific, the proposed method consists of three parts: i) hierarchical code partitioning; ii) pre-processing; iii) soft-output wMD decoding. During a coherence time, the part i) is performed at once in channel training phase while the parts ii) and iii) are performed at each time slot in data transmission phase (see Fig.~\ref{channel_training}). The detailed procedures are described as follows.

\subsection{Channel Training Phase}\label{subsec:CT}

In this phase, the BS first estimates a channel matrix  $\hat{\Hm}$ using the $T_{\rm t}$ pilot signals where numerous channel estimation methods can be used (see \cite{Choi} and \cite{Li} for details). Using the $\hat{\Hm}$, the BS creates the 
 spatial-domain code $\Cc=\{\cv_{0},...,\cv_{m^K-1}\}$, defined in (\ref{eq:code-def}), and computes the weights (channel reliabilities) of $N$ parallel channels  $\alphav_{\ell}=(\alpha_{\ell,1},...,\alpha_{\ell,N})$. Then, the (soft-output) wMD decoding can be performed.

\begin{figure}
\centerline{\includegraphics[width=9cm]{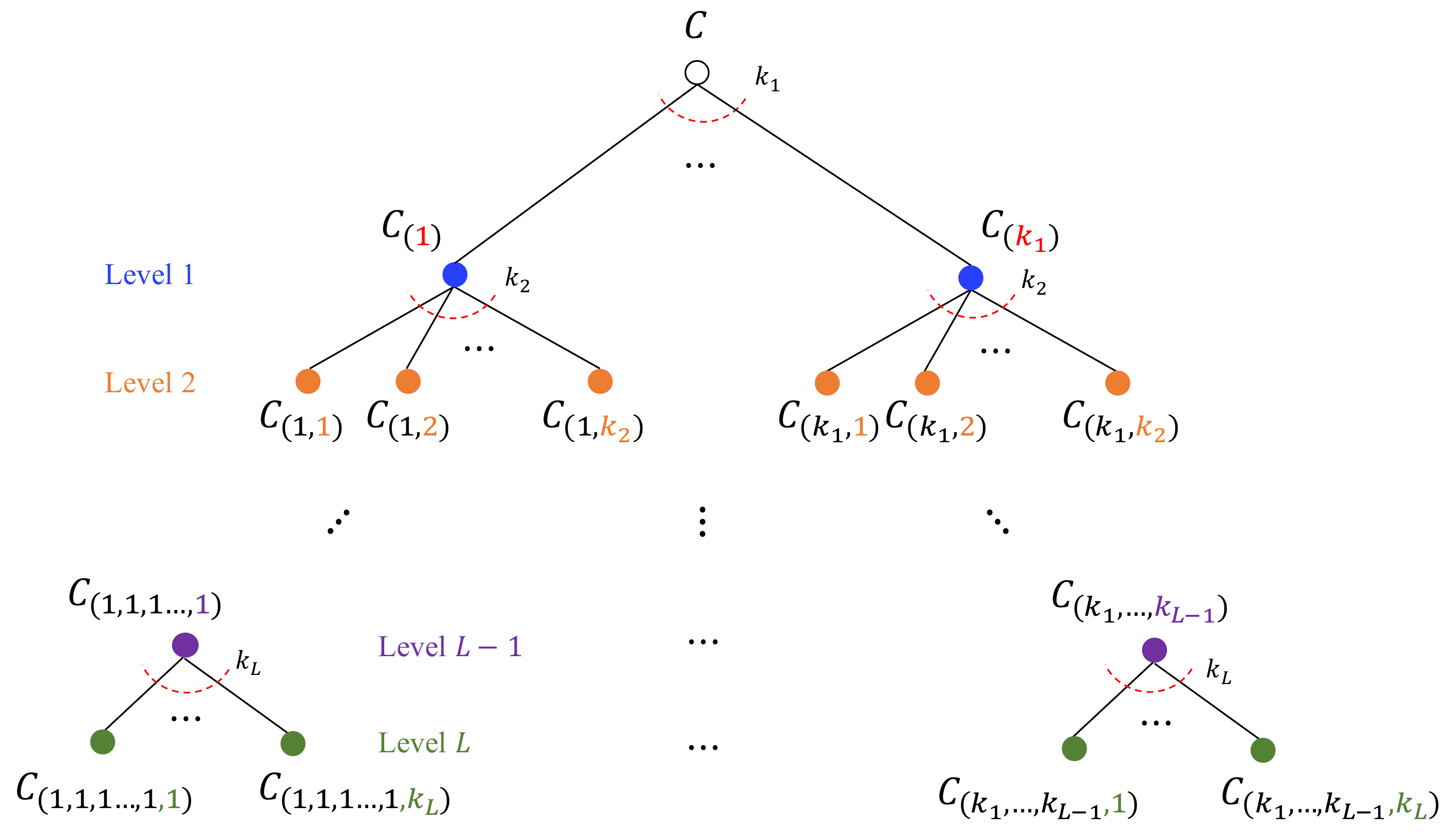}}
\caption{Hierarchical code partitioning.}
\label{multilevel}
\end{figure}

The following procedures are required to perform the low-complexity (soft-output) wMD decoding. For a fixed hierarchical level $L \geq 1$, the code $\Cc$ is partitioned into several subcodes in a hierarchical manner:

\begin{itemize}
\item At the level-1, using a vector quantization method, the $\Cc$ is partitioned into the $k_1$ subcodes $\Cc_{(1)},...,\Cc_{(k_1)}$ with $\cup_{i=1}^{k_1}\Cc_{(i)} = \Cc$. In this paper, as the vector quantization method, we use the $k$-means clustering algorithm in \cite{Lloyd} with Hamming distance metric. Also, this algorithm generates the $k_1$  {\em centroids} $\{\muv_{(i)}: i=1,2,...,k_1\}$, where each $\muv_{(i)}$ is a length-$N$ binary vector. For each centroid $\muv_{(i)}=(\mu_{(i)}^{1},...,\mu_{(i)}^{N})$, the weight vector $\betav_{(i)}=(\beta_{(i)}^{1},...,\beta_{(i)}^{N})$ is computed as
\begin{equation}
\beta_{(i)}^{j} =- \log{ \frac{1}{|\Cc_{(i)}|} \sum_{\cv \in \Cc_{(i)}} d_{\rm h}(c_j, \mu_{(i)}^{j})}, \label{eq:partition-weight}
\end{equation}for $j=1,...,N$, where $d_{\rm h}(\cdot,\cdot)$ denotes the Hamming distance. As in wMD decoding, the purpose of such weights is to allocate a higher belief to the locations having more dominant occurrences.

\item At the level-2, each level-1 subcode $\Cc_{(i_1)}$ is further partitioned into the  $k_2$  subcodes $\Cc_{(i_1,i)}$ for $i=1,...,k_2$ using the $k$-means clustering algorithm. They satisfy the
\begin{equation}
\bigcup_{i=1}^{k_2}\Cc_{(i_1,i)} = \Cc_{(i_1)}.
\end{equation}
Also, the $k_2$ centroids $\{\muv_{(i_1,i)}:i=1,2,...,k_2\}$ are generated and for each centroid $\muv_{(i_1,i)}$, the weight vector $\betav_{(i_1,i)}$ is computed using (\ref{eq:partition-weight}).

\item Generally at the level-$\ell$, each level-$(\ell-1)$ subcode $\Cc_{(i_1,i_2,...,i_{\ell-1})}$ is further partitioned into the $k_{\ell}$ subcodes $\Cc_{(i_1,i_2,...,i_{\ell-1},i)}$ for $i=1,...,k_{\ell}$, and the corresponding $k_{\ell}$ centroids $\{\muv_{(i_1,...,i_{\ell-1},i)}:i=1,2,...,k_{\ell}\}$ are generated. Also, for each centroid $\muv_{(i_1,...,i_{\ell-1},i)}$, the weight vector  $\betav_{(i_1,...,i_{\ell-1},i)}$ is computed using the (\ref{eq:partition-weight}).

\item Repeatedly perform the above process for $\ell=1,...,L$.
\end{itemize} The above process is referred to as {\em hierarchical code partitioning} because this process partitions the code $\Cc$ into the subcodes with the hierarchical structure (see Fig.~\ref{multilevel}). Note that the resulting subcodes are used during the coherence time (e.g., $T$ time slots), as shown in Fig.~\ref{channel_training}.


\subsection{Data Transmission Phase}\label{subsec:DT}

In the data transmission, the decoding consists of the two parts as pre-processing and (soft-output) wMD decoding. In the pre-processing, some unnecessary codewords (having a lower probability to be a valid codeword) are precluded, and then the (soft-output) wMD decoding is performed using the reduced code.

\begin{figure}
\centerline{\includegraphics[width=8cm]{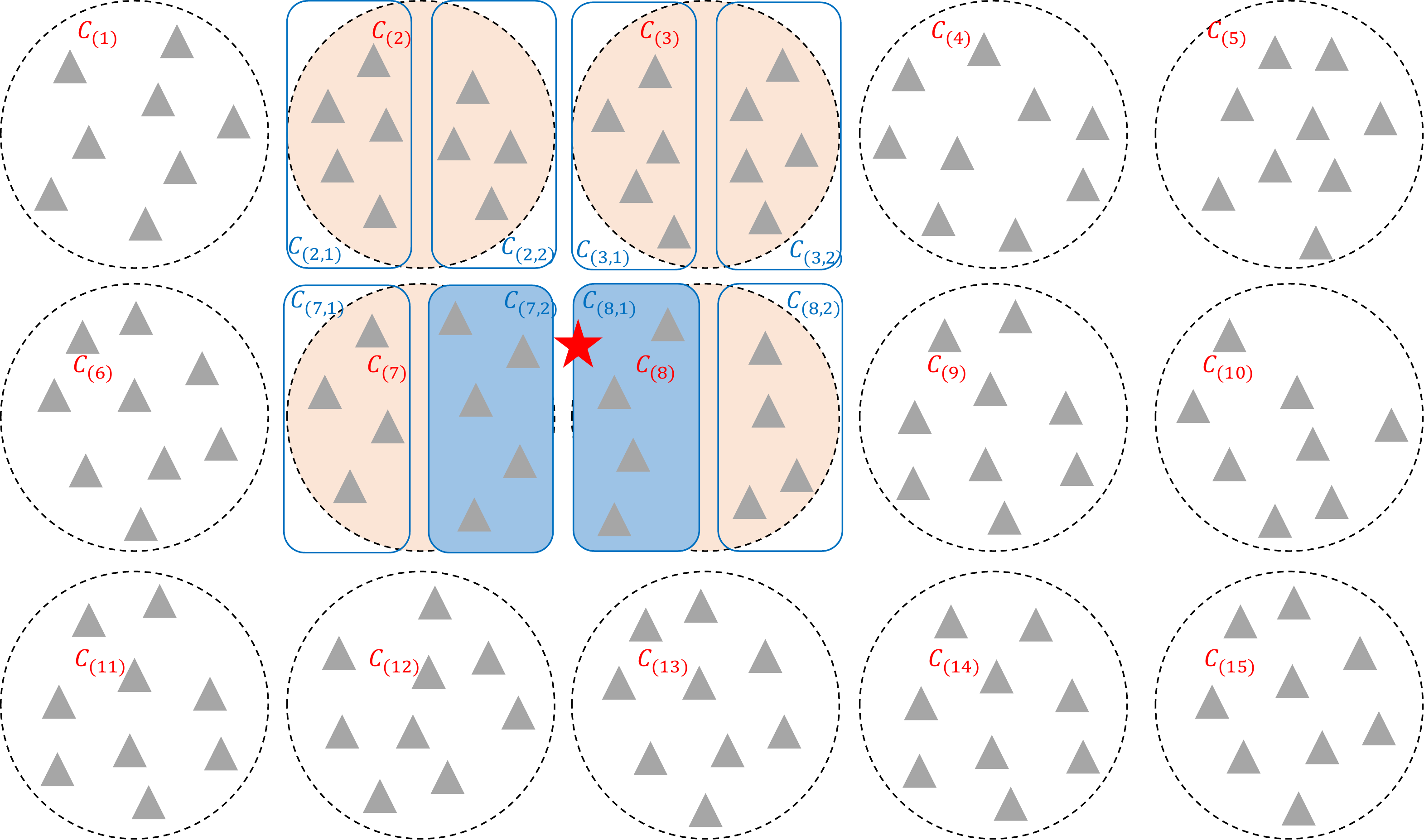}}
\caption{Illustration of the pre-processing when 2-level hierarchical code partitioning is used.  The triangle nodes denote the codewords of the $\Cc$ and the star node denotes the received observation. Also, the dashed circles denote the level-1 subcodes and the solid squares denote the level-2 subcodes. After the pre-processing, the black-colored triangle nodes are only remained in the search-space.}
\label{cluster}
\end{figure}

{\bf 1) Pre-processing:}  As shown in Fig.~\ref{cluster}, this process is performed as follows.
\begin{itemize}
\item With the weight vector $\betav_{(i)}$, the weighted Hamming distances between the $\rv[t]$ and the level-1 centroids $\muv_{(1)},...,\muv_{(k_1)}$ is are computed as

\begin{equation}
d_{i} = d_{\rm wh} (\muv_{(i)}, \rv[t], \betav_{(i)}), \label{eq:dist_centroid}
\end{equation} for $i=1,...,k_1$.  Sort the $d_i$'s in an increasing order and then define the index set containing the first $q_1$ indices as $\Ic_{1}=\{i_1,i_2,...,i_{q_1}\}$. In this process, the codewords outside the chosen subcodes are  eliminated from the search-space.

\item Similarly, with the weight vectors $\{\betav_{(i_1,i)}: i_{1} \in \Ic_{1}, i =1,...,k_2\}$, the weighted Hamming distances between the $\rv[t]$ and the level-2 centroids $\{\muv_{(i_1,i)}: i_{1} \in \Ic_{1}, i =1,...,k_2\}$ are computed, and then 
the corresponding index set $\Ic_{2}=\{(i_1,i_2)\}$ with $|\Ic_{2}|=q_{2}$ is defined. Note that this process further reduces the search-space by ruling out the unnecessary codewords.

 \item In general, the weighted Hamming distances between the $\rv[t]$ and the level-$\ell$ centroids $\{\muv_{(i_1,...,i_{\ell-1},i)}: (i_1,...,i_{\ell-1}) \in \Ic_{\ell-1}, i=1,...,k_{\ell}\}$ are computed with the weight vectors $\{\betav_{(i_1,...,i_{\ell-1},i)}: (i_1,...,i_{\ell-1}) \in \Ic_{\ell-1}, i=1,...,k_{\ell}\}$, and the corresponding index set $\Ic_{\ell}$ with $|\Ic_{\ell}|=q_{\ell}$ is defined.
 \item Repeatedly perform the above process for $\ell=1,2,...,L$.
 \end{itemize} From the pre-processing, the reduced code $\Cc_{\rm r}(\rv[t]) \subset \Cc$ is obtained as
 \begin{equation}
 \Cc_{\rm r}(\rv[t]) = \bigcup_{(i_1,i_2,...,i_L) \in \Ic_{L}} \Cc_{(i_1,i_2,...,i_L)}.
 \end{equation} Note that the $\Cc_{\rm r}(\rv[t])$ depends on the current observation $\rv[t]$ and only contains the codewords which are close to the $\rv[t]$ in some sense.  It is noticeable that in the proposed method, the $q_{\ell}>1$ subcodes can be chosen concurrently for each level $\ell$. This is to improve the probability that a valid codeword belongs to the $\Cc_{\rm r}(\rv[t])$, with the expense of the complexity.  Therefore, the parameters $\{(k_1,...,k_L), (q_1,...,q_L)\}$ should be carefully chosen by taking the performance-complexity tradeoff into account. Also, since he number of chosen subcodes at the level $\ell$ should be smaller than the remaining subcodes at the level $\ell-1$, the parameters should satisfy the condition of 
 \begin{equation}
q_{\ell} \leq q_{\ell-1}k_{\ell},
\end{equation} for $\ell=1,...,L$, where $q_0 = 1$.

\vspace{0.1cm}
{\bf 2) (soft-output) wMD decoding:} The wMD decoding with either hard-outputs or soft-outputs is performed with the reduced code $\Cc_{\rm}(\rv[t])$ for each time slot $t$.

\vspace{0.1cm} 
\begin{example} Fig.~\ref{cluster} shows the hierarchical code structure and the pre-processing for $L=2$, where the triangles denote the codewords of the $\Cc$ and the star denotes the received observation $\rv$. In this example, the code $\Cc$ is partitioned into the 15 subcodes (represented by the circles in Fig.~\ref{cluster}) and each level-1 subcode is further partitioned into the 2 subcodes (represented by the squares in Fig.~\ref{cluster}). Also, the pre-processing can be explained as follows. At the level-1, the 4 subcodes $\Cc_{(2)}, \Cc_{(3)}, \Cc_{(7)}, \Cc_{(8)}$  (denoted by the filled circles) are chosen and then at the level-2, the 2 subcodes $\Cc_{(7,2)}, \Cc_{(8,1)}$ (denoted by the filled squares) are chosen. After this process, the wMD decoding is performed with the codewords belong to the $\Cc_{(7,2)}\cup \Cc_{(8,1)}$.
\end{example}
\vspace{0.2cm}

\begin{figure}[t]
\centerline{\includegraphics[width=9cm]{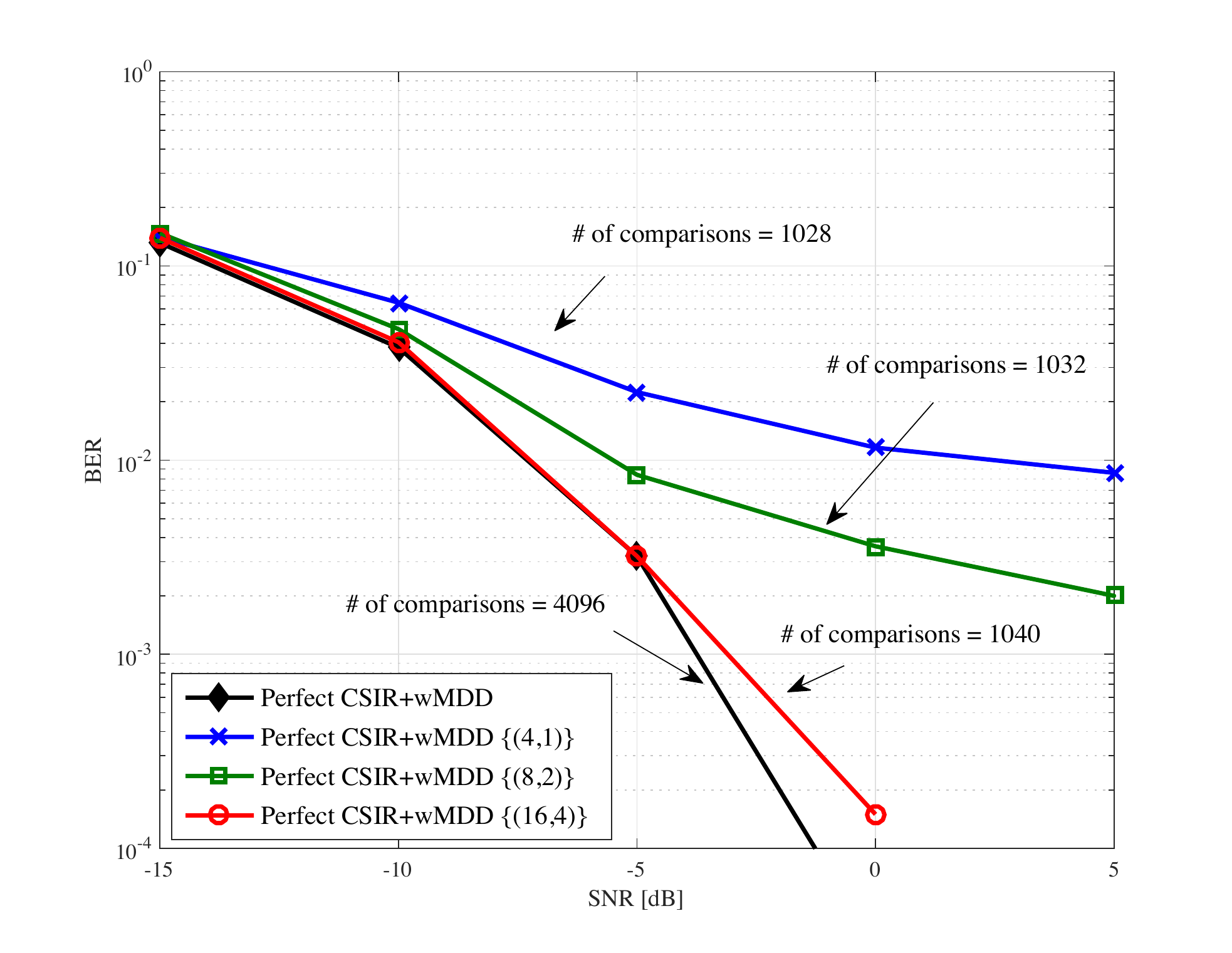}}\vspace{-0.5cm}
\caption{$K=6$ and $N_{\rm r}=64$. BER performances of the proposed method according to the choices of $\{(k_1,q_1)\}$ when 1-hierarchical level is considered.}
\label{comp1}
\end{figure}

\subsection{Discussion on Computational Complexity}

In this section, we discuss the complexity of the low-complexity wMD decoding for each coherence time $T_{\rm c}=T_{\rm t}+T_{\rm d}$, where the complexity is measured as the number of distance comparisons. Let $\Nc_{\rm cp}$, $\Nc_{\rm pre}$, and $\Nc_{\rm wMD}$ denote the number of distance comparisons required for hierarchical code partitioning, pre-processing, and wMD decoding, respectively. Then, the overall complexity during the coherence time $T_{\rm c}$ is given by $\Nc_{\rm cp} + T_{\rm d}(\Nc_{\rm pre}+\Nc_{\rm wMD})$. Accordingly, the average complexity per time slot is given by
\begin{align}
\Nc_{{\rm total}} &= \frac{1}{T_{\rm c}} \Nc_{\rm cp}+  \frac{T_{\rm d}}{T_{\rm c}} \left(\Nc_{\rm pre}+\Nc_{\rm wMD}\right)\\
&\approx  \Nc_{\rm pre}+\Nc_{\rm wMD},
\end{align} where the above approximation is generally accurate since $T_{\rm d} \gg T_{\rm t}$ and $T_{\rm c} \gg 1$.  Thus, we assume the $\Nc_{\rm total}=\Nc_{\rm pre}+\Nc_{\rm wMD}$ as the average complexity per time slot.

First, the pre-processing complexity is computed as
\begin{equation}
\Nc_{{\rm pre}} = \sum_{\ell=1}^{L} q_{\ell-1}k_{\ell},
\end{equation} where $q_0 = 1$, since there are the $q_{\ell-1}k_{\ell}$ number of centroids for each level $\ell$.  After the pre-processing, the number of the remaining codewords in the search-space is 
\begin{equation}
\Nc_{{\rm wMD}}=\sum_{(i_1,...,i_L)\in \Ic_{L}} |\Cc_{(i_1,...,i_L)}|.
\end{equation} In fact, the $\Nc_{{\rm wMD}}$ is not a constant but is determined as a function of  a channel matrix $\Hm$ and an observation $\rv[t]$. This is because the $k$-means clustering algorithm does not ensure the equi-partitioning of the code \cite{Lloyd}. Via numerical results, we verified that the average value of $\Nc_{{\rm wMD}}$, where the average is performed over a random channel matrix, is very well approximated to the complexity obtained with the assumption of the uniform partitioning as
\begin{align*}
\Nc_{{\rm wMD}}&\approx |\Wc|^K\times \frac{q_1}{k_1}\times \frac{q_2}{q_1 k_2}\cdots\times \frac{q_L}{q_{L-1} k_L}\\
&=|\Wc|^K\frac{q_L}{\prod_{\ell=1}^{L}k_{\ell}}.
\end{align*} With this approximation, the average decoding complexity per time slot is given by
\begin{align}
\Nc_{\rm total} &= \Nc_{{\rm pre}}+\Nc_{{\rm wMD}}\\
& \approx \sum_{\ell=1}^{L} q_{\ell-1}k_{\ell} + |\Wc|^K\frac{q_L}{\prod_{\ell=1}^{L}k_{\ell}}, \label{eq:complexity}
\end{align} which is assumed as the average complexity  in the sequel.

\vspace{0.2cm} 
\begin{example} Consider the uplink MIMO systems with $K=8$ and $N_{\rm r} = 64$ where 4-QAM is assumed. The overall complexity of wMD decoding is very expensive as $\Nc_{\rm total} \approx 65536$. Using the 1-hierarchical level $\{(32,8)\}$, the complexity can be reduced to the $25\%$ of the original complexity as $\Nc_{\rm total} \approx 16416$. Also, using the 3-hierarchical level $\{(32,4,4),(8,8,8)\}$, the complexity can be further reduced to the $1.7\%$ of the original complexity as $\Nc_{\rm total} \approx 1120$. In Fig.~\ref{comp2}, it is shown that the performance obtained with the 3-hierarchical level approaches the optimal performance of the wMD decoding. 
\end{example}

\section{Numerical Results}\label{sec:numerical}

We evaluate the performances of the low-complexity (soft-output) wMD decoding. A Rayleigh fading channel is assumed in which each element of a channel matrix $\Hm$ is drawn from an independent and identically distributed (i.i.d.) circularly symmetric complex Gaussian random variable with zero mean and unit variance. Also, 4-QAM and ZF-type channel estimation method in \cite{Choi} are assumed.

\begin{figure}[t]
\centerline{\includegraphics[width=9cm]{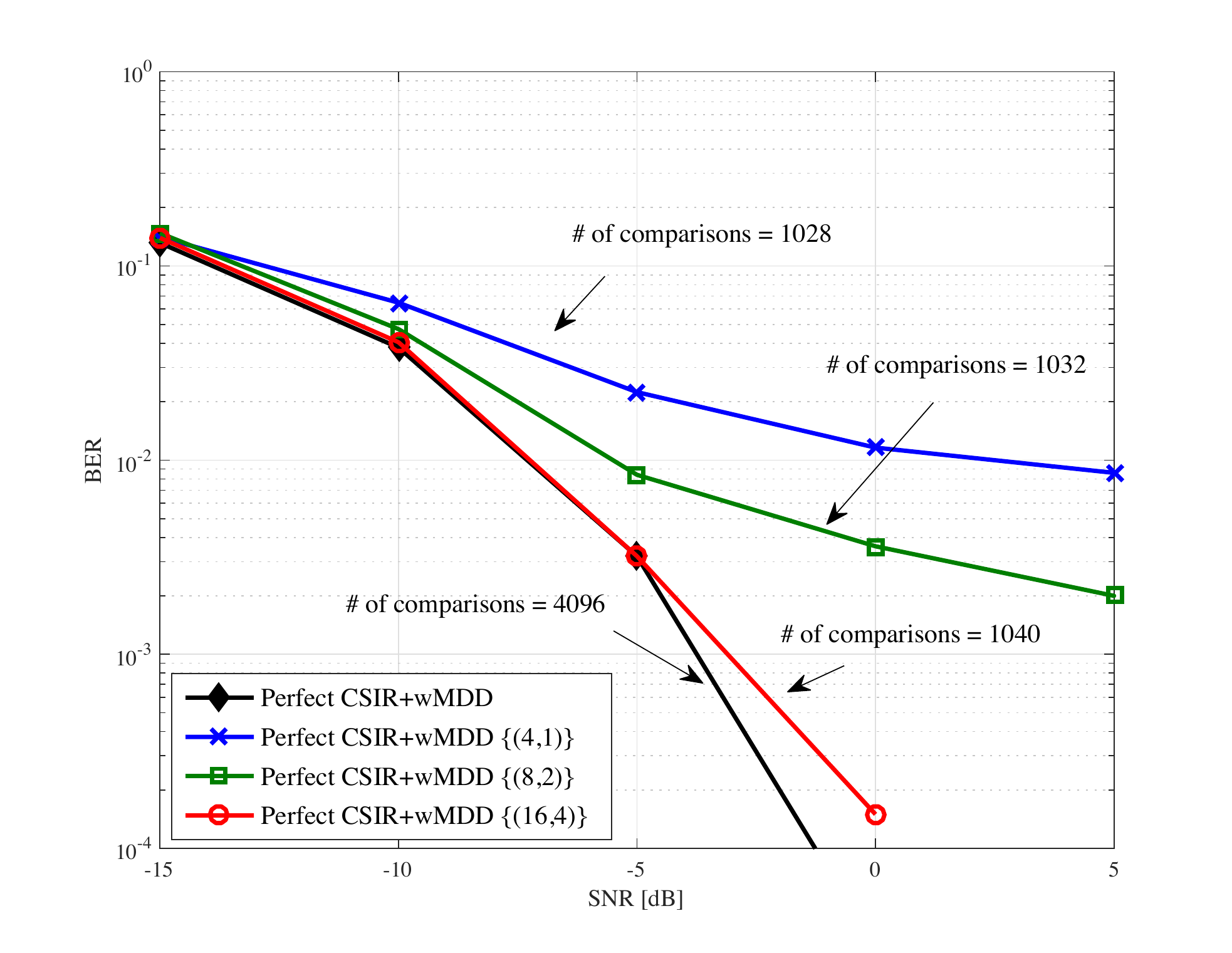}}\vspace{-0.5cm}
\caption{$K=6$ and $N_{\rm r}=64$. BER performances of the proposed method according to the choices of $\{(k_1,q_1)\}$ when 1-hierarchical level is considered.}
\label{comp1}
\end{figure}

\begin{figure}[t]
\centerline{\includegraphics[width=9cm]{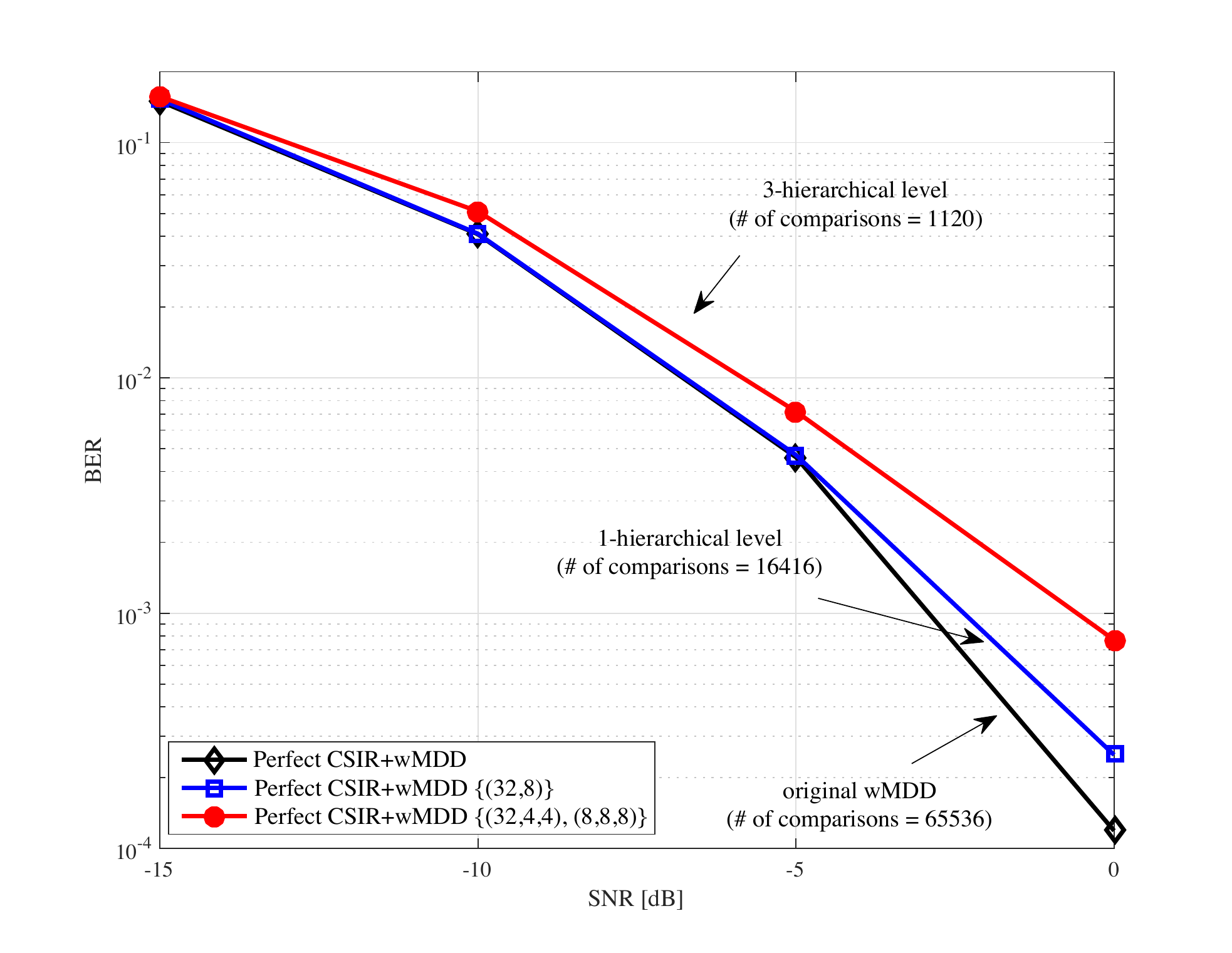}}\vspace{-0.5cm}
\caption{$K=8$ and $N_{\rm r}=64$. Uncoded BER performances of the proposed method according to the number of hierarchical levels.}
\label{comp2}
\end{figure}

Fig.~\ref{comp1} shows the BER performances of the low-complexity wMD decoding according to the choices of  $\{(k_1,q_1)\}$, where the parameters are chosen such that the size of the reduced code (i.e., search-space) is equal to 1024. We observe that the BER performance is enhanced by partitioning the code $\Cc$ into the subcodes with a smaller size, and the performance gain is unbounded due to the error-floor. From this observation, the best strategy for for selecting the $\{(k_1,q_1)\}$ is to choose a larger $k_1$ as long as the complexity of the pre-processing is relatively small compared to the complexity of the wMD decoding.

Fig.~\ref{comp2} shows the BER performances of the low-complexity wMD decoding as a function of a hierarchical level $L$. This example shows that, using the 3-hierarchical level, the complexity is significantly reduced to the $1.7\%$ of the original complexity with a negligible performance loss. Hence, it is expected that the use of a larger hierarchical level is beneficial as $K$ increases.



\begin{figure}
\centerline{\includegraphics[width=9cm]{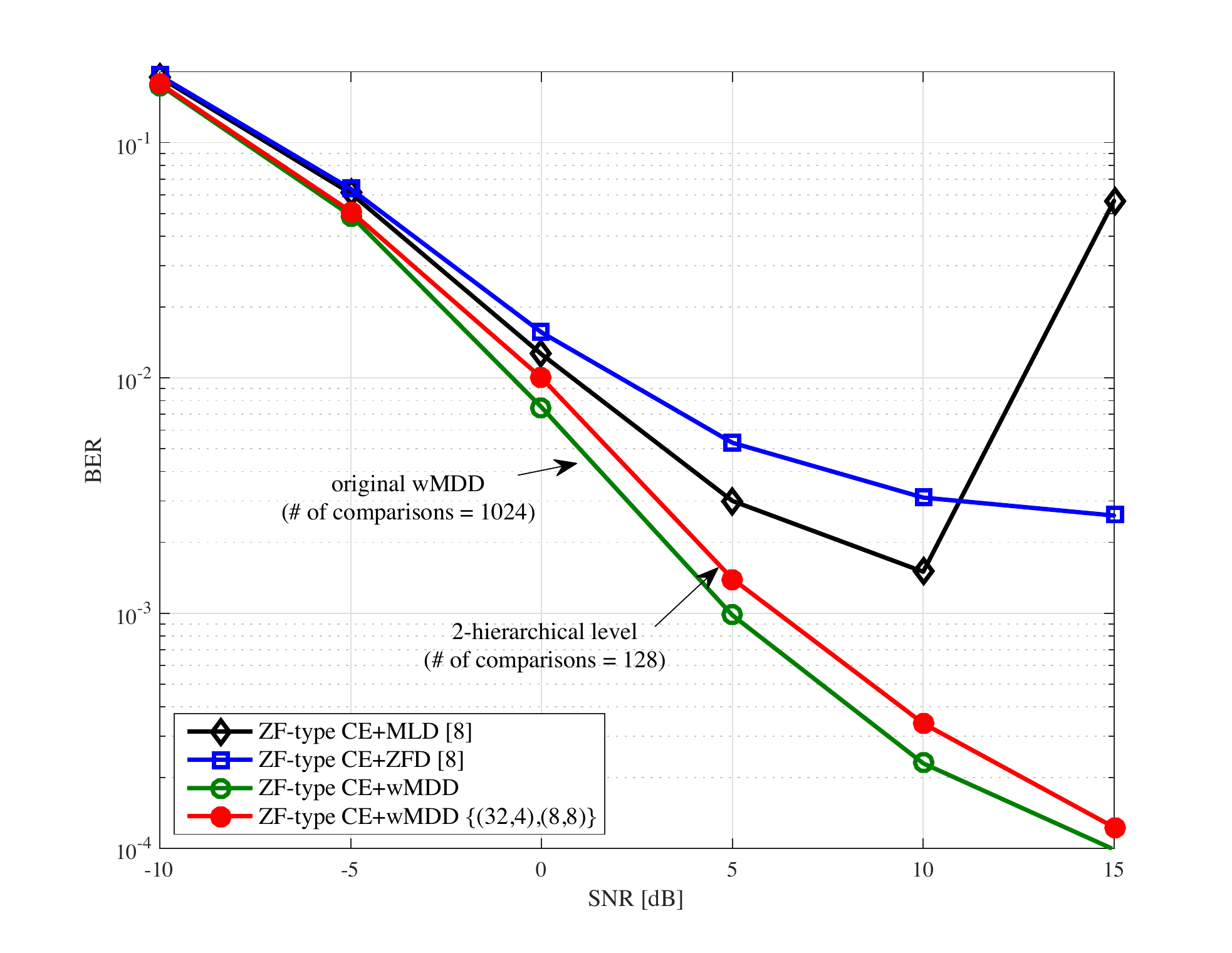}}\vspace{-0.5cm}
\caption{$K=5$ and $N_{\rm r}=32$. Performance comparisons of the various MIMO detection methods. The training overhead is set to $T_{\rm t}=25$.}
\label{comp3}
\end{figure}

In Fig.~\ref{comp3}, we compare the low-complexity wMD decoding with the existing MIMO detection methods. A block fading duration (i.e., coherence time) is set to be $T_{\rm c} = T_{\rm t} + T_{\rm d} = 1000$ time slots and the training overhead is set to the $2.5\%$ of the coherence time (i.e., $T_{\rm t} = 25$). For the comparisons, we consider the ML and ZF detection methods in \cite{Choi}. It is noticeable that the ML detection with imperfect CSIR severely suffers from the BER degradation especially in the high-SNR regimes due to the impact of the inaccurate CSIR. In contrast, the wMD decoding with imperfect CSIR outperforms the existing MIMO detection techniques and the performance gaps increase as $\SNR$ grows. Namely, the wMD decoding is more robust to imperfect CSIR than ML detection although both methods achieve the same optimal performance with perfect CSIR. We notice that the use of 2-hierarchical level can reduce the decoding complexity of the $10\%$ of the original complexity with a small performance loss. Thus, in an imperfect CSIR, the low-complexity wMD decoding can provide a satisfactory performance with a manageable decoding complexity.


\begin{figure}
\centerline{\includegraphics[width=9cm]{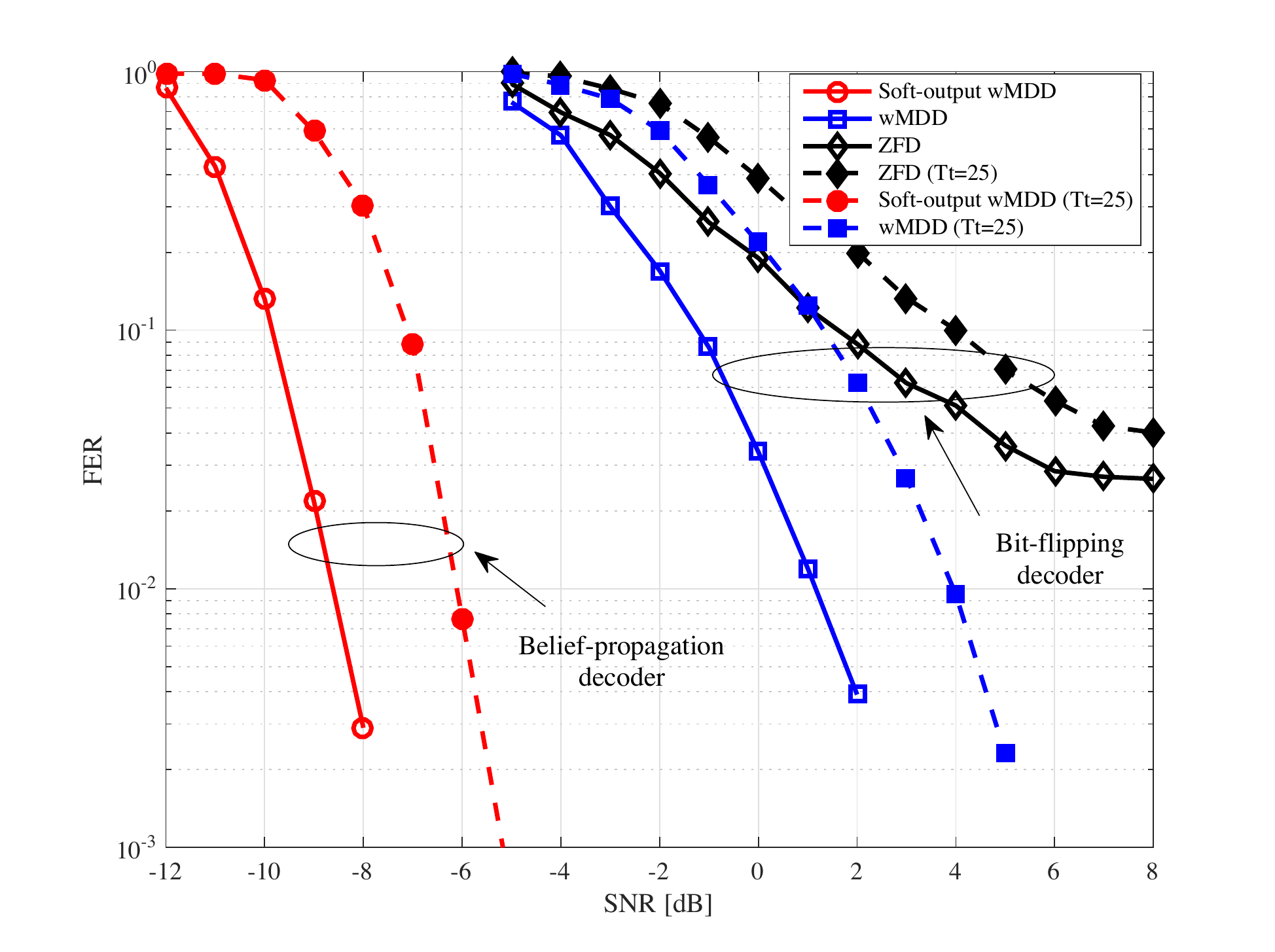}}\vspace{-0.5cm}
\caption{$K=5$ and $N_{\rm r}=32$. Performance comparisons of the various MIMO detection method for the coded MIMO system. A perfect CSIR is assumed for the solid lines and ZF-type channel estimation with $T_{\rm t}=25$ is assumed for the dashed lines.}
\label{coded_FER}
\end{figure}

Fig.~\ref{coded_FER} shows the coded frame-error rate (FER) performances of the various MIMO detection methods where the coded system is formed by concatenating a MIMO detector with a low-density-parity-check (LDPC) code. We adopt a rate 1/2 LDPC code of the blocklength 672 from the IEEE802.11ad standardization \cite{IEEE}. As an LDPC code decoder, the bit-flipping decoder \cite{Rao} is used for the wMD decoding and ZF-type detector, and the belief-propagation decoder \cite{Richardson} is used for the soft-output wMD decoding. Also, to reduce the complexity of the (soft-output) wMD decoding, the 2-hierarchical level  $\{(32,4),(8,8)\}$ is assumed. In this example, a block fading duration (i.e., coherence time) is set to be $T_{\rm c} = T_{\rm t} + T_{\rm d} = 1369$ with $T_{\rm t} = 25$ and $T_{\rm d} = 1274$, where the two coded outputs of the LDPC code are transmitted during the coherence time. This example shows that the soft-output wMD decoding has a non-trivial performance gain over the wMD decoding (or ML detector) and ZF-type detector with a comparable complexity.

\section{Conclusion}\label{sec:conclusion}

We proposed the soft-output wMD decoding which efficiently computes the soft outputs (e.g., log-likelihood ratios) from one-bit quantized observations. This enables to employ {\em soft} channel decoder (e.g., belief-propagation decoder) for the MIMO systems with one-bit ADCs. Furthermore, we presented the low-complexity soft-output wMD decoding by introducing hierarchical code partitioning, which can be regarded as a sphere decoding for the MIMO systems with one-bit ADCs. Finally we demonstrated that the proposed method significantly outperforms the other MIMO detectors with hard-decision outputs, with a comparable complexity. One possible extension is to study the soft-output wMD decoding for a slowly varying channel, in which we may reduce the channel training overhead by updating the spatial-domain code and the weights from the previous ones, rather than newly constructing them.






\end{document}